\RequirePackage{fix-cm}
\documentclass[smallextended,natbib]{svjour3}       
\smartqed  
\usepackage{graphicx}

\begin{document}

\title{Research Performance of Turkish Astronomers in the Period of 1980$-$2010
}


\author{Sel{\c c}uk Bilir              \and
        Ersin G{\"o}{\u g}{\"u}{\c s}  \and
       {\"O}zgecan {\"O}nal            \and
        Nazl\i~Derya {\"O}zt{\"u}rkmen \and
        Talar Yontan
}


\institute{S. Bilir \at
Istanbul University, Faculty of Science, Department 
of Astronomy and Space Sciences, 34119, University, Istanbul, Turkey\\ 
              Tel.: +90-212-440 00 00 - 10534\\
              Fax:  +90-212-440 03 70\\
              \email{sbilir@istanbul.edu.tr}\\
           \and
           E. G{\"o}{\u g}{\"u}{\c s} \at
Sabanc\i~University, Faculty of Engineering and Natural
  Sciences, 34956, Orhanl\i-Tuzla, Istanbul, Turkey\\
           \and
           {\"O}. {\"O}nal  \at
Istanbul University, Faculty of Science, Department 
of Astronomy and Space Sciences, 34119, University, Istanbul, Turkey\\ 
           \and 
           N. D. {\"O}zt{\"u}rkmen \at
Istanbul University, Faculty of Science, Department 
of Astronomy and Space Sciences, 34119, University, Istanbul, Turkey\\ 
           \and 
           T. Yontan \at
Istanbul University, Faculty of Science, Department 
of Astronomy and Space Sciences, 34119, University, Istanbul, Turkey\\ 
}

\date{Received: date / Accepted: date}

\maketitle

\begin{abstract}

We investigated the development of astronomy and astrophysics research 
productivity in Turkey in terms of publication output and their impacts 
as reflected in the Science Citation Index (SCI) for the period 1980-2010. 
Our study involves 838 refereed publications, including 801 articles, 16 letters, 
15 reviews, and six research notes. The number of papers were prominently 
increased after 2000 and the average number of papers per researcher is 
calculated as 0.89. Total number of received citations for 838 papers is 6938, 
while number of citations per papers is approximately 8.3 in 30 years. 
Publication performance of Turkish astronomers and astrophysicists was 
compared with those of seven countries that have similar gross domestic 
expenditures on research and development, and members of Organization for 
Economic Co-operation and Development (OECD). Our study reveals that the 
output of astronomy and astrophysics research in Turkey has gradually 
increased over the years.

\end{abstract}

\section{Introduction}
The 1933 university reform marks an important cornerstone for the development of 
contemporary teaching and scientific research in Turkey. Even though the roots 
of astronomy within the present geographical territories of Turkey dates back 
to antiquity (e.g., Hipparchos of Nicaea), modern astronomy research in Turkey, 
along with many other fields in science, gained its founding principles with the 
1933 university reform. A rather rapid progress was made in the subsequent years 
mainly with the help of visiting German scientists who escaped the suppression and 
persecution of the regime in Nazi Germany \citep{Ishakoglu98}. By 1960, there were 
already 48 papers published in refereed international journals at three well 
established astronomy institutes \citep{Ozkan04, Inonu09}.

1970s and onward has witnessed rapid changes in astronomy research infrastructure. 
With the 1973 Universities Act, astronomical institutes were transformed into 
departments. Following the regulation change by the Higher Education Council in 
1982, the departments were named Astronomy and Space Sciences to carry on both 
undergraduate and graduate teaching, as well as performing astronomical research 
\citep{Eker12}. In the meantime, the necessity for larger telescopes arose, which 
sparked the idea of a national observatory in 1960s. After very intensive astronomical 
testing of numerous sites by Turkish astronomers, Antalya-Bak\i rl\i tepe was 
decided to establish the national observatory \citep{Aslan89}. The national 
observatory became operational in 1997 and has constantly been improving in 
instrumental capabilities since then. Over the recent years, few research groups 
embraced into astrophysical investigations with multi-wavelength approach rather 
than being limited to the ground based optical window. This scheme provides 
important opportunities for Turkish astronomers and astrophysicists to engage 
into international collaborations.

Publication performance of researchers based in Turkey and impact of those 
publications have been studied for time periods of about 10 years: \cite{Demircan88} 
used the Science Index and Citation Index to determine the performance statistics 
of Turkish researchers in the time span from 1975 to 1984. \cite{Derman92} expanded 
this study by searching additional five years (until 1989). A similar investigation 
was performed by \cite{Uzun96} covering the period of 1985-1994. \cite{Ozkan04} 
explored institutional developments and growth in the variety of study fields in 
Turkish astronomy from the foundation of the Turkish Republic in 1923 until 2003. 
\cite{Demircan08} provided the performance profiles of all researchers in Turkey for 
the years of 2006 and 2007. \cite{Inonu09} presents one of the most extensive 
panorama of astronomy and astrophysics studies in Turkey, including historical 
developments of main astronomy departments. Finally, present day astronomy and its 
development in Turkey were studied by \cite{Eker12}.

Assessing collective performance of scientific research has been a complicated 
issue. \cite{Martin83} investigated indicators of scientific progress via numerous 
channels, including citation based indicators of the impact of research 
articles in radio astronomy. Some variants of citation related indicators include 
simply the number of citations, citations per publication and peak year citations 
per publication \citep{Hsieh04}. More recently, the h-index \citep{Hirsch05} has 
been instrumental in assessing the research output of an individual, as well as 
research groups.

In this paper, a thorough investigation of all publications in the fields of 
astronomy and astrophysics between 1980 and 2010, which are in the Science 
Citation Index (SCI) and includes researchers based in Turkey, has been performed. 
Some indicative results have been compared with those of OECD countries which 
have similar gross domestic expenditure. The method employed to determine the 
research performance in the investigated 30 years is presented in Section 2. 
Results of these investigations and discussion of the main results are 
presented in Sections 3 and 4, respectively.

\section{Method}
Thomson Reuters Web of Knowledge\footnote{http://apps.webofknowledge.com}, 
which includes twelve different databases, is used while searching papers in 
astronomy and astrophysics. The database provides a list of SCI expanded 
journals and citation records of papers since 1980 to present day. 
All 55 refereed journals with SCI subject category of ``Astronomy \& 
Astrophysics'' were identified for the thirty year period. Note that such a
subject field delineation scheme would exclude articles published in 
multidisciplinary journals \citep{Rinia93}. However, the number of 
papers published in such journals originating from Turkey between 1980 
and 2010 is very few and our applied scheme would still be a reliable indicator.

The search for papers by authors or co-authors based in Turkey between 1980-2010 
resulted in 1394 papers. These papers were found in astronomy and astrophysics 
related fields but also included sub-branches of molecular physics, meteorology 
and atmospheric sciences, geochemistry, geophysics, space engineering, multi 
disciplinary math applications, multi disciplinary physics, and astrobiology. 
It was, therefore, needed to make a refinement by investigating the keywords
and abstracts of each individual article to include only papers in 
astronomy and astrophysics. As a result, 886 papers were reached. Publication 
type is another criteria which was considered next. It has been seen that these 
886 papers had been divided into six groups based on their types: articles (801), 
proceedings (37), letters (16), reviews (15), errata (10), and research notes (7). 
It was also found that some of these studies were presented at meetings 
before they have been published, some of them were corrected then re-published 
and {one research note was the summary of a dissertation}. Because of these reasons only 
the articles, letters, reviews, and scientifically motivated research notes were 
considered in this study. These conditions resulted in 838 publications. 

\begin{figure}
\centering
   \includegraphics[scale=0.55]{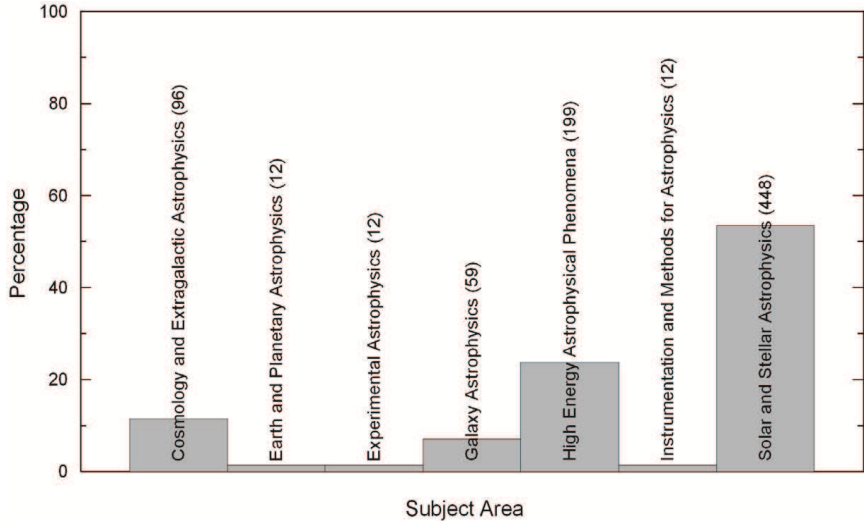}
   \caption{The distribution of 838 papers over subject areas. The numbers in parenthesis 
    show that the number of papers belong to relevant subject area.}
\label{fields}
\end{figure}

The subject classification scheme of the arXiv database\footnote{http://www.arxiv.org} 
of the Cornell University Library, is used while categorizing these 838 publications 
according to their subjects in astronomy and astrophysics
\footnote{The subject classification of the arXiv database was adopted since
it is an open access resource and used heavily by researchers. This classification
scheme was used commonly in the literature \citep[e.g.][]{Cho08}}. These categories are 
cosmology and extragalactic astrophysics, Earth and planetary astrophysics, galaxy 
astrophysics, high energy astrophysical phenomena, instrumentation and methods for 
astrophysics, solar and stellar astrophysics. These 838 papers were divided into 
above categories by considering their keywords and abstracts. Distribution of papers 
based on their subjects areas were presented in Fig. 1. It is clear that the most of 
the papers were published in solar and stellar astrophysics (53.5\%). This is followed 
by high energy astrophysical phenomena (23.8\%), cosmology and extragalactic 
astrophysics (11.5\%) and galaxy astrophysics (7.0\%). These four subject categories 
constitute about 96\% of the subjects of publications studied in Turkey. 

\section{Results}
We have obtained 838 astronomy and astrophysics papers that have been published 
in SCI journals between 1980 and 2010 based on the abovementioned criteria. We 
have analyzed these papers according to their release years, number of citations, 
and collaboration countries. Furthermore, we proceeded our analysis with considering 
the academic levels of the authors, the subject areas and annual impact factors of 
the journals. Finally, we compared these results with some OECD countries having 
similar gross domestic expenditures on research and development.

\subsection{The Distribution of Annual Number of Papers and their Citations}
The distribution of the papers published by researchers from Turkey in SCI journals 
is shown in Fig. 2. There are two cases shown in this figure; one is for all papers 
(Fig. 2a) and the other one is for papers with the corresponding authors from Turkey 
(Fig. 2b). 

\begin{figure}
\centering
   \includegraphics[scale=0.30]{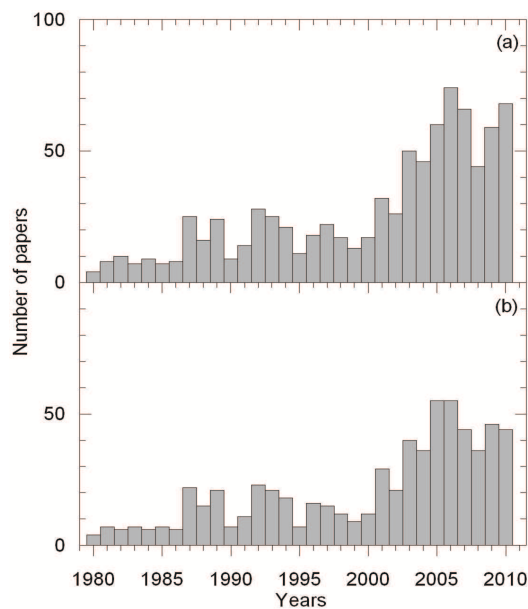}
   \caption{Distribution of papers that published in SCI journals between 1980-2010 
   in years: (a) all the papers, (b) the corresponding authors from Turkey.}
\label{fields}
\end{figure}

It is clear from Fig. 2a that there exists three prominent periods; 1980-1987, 1988-2000, 
and 2001-2010. The average number of papers annually in these periods are 9.8, 18.1, and 
52.5, respectively. If we take into account the papers with corresponding authors from 
Turkey, no obvious overall difference in all three time periods were seen and the average 
annual number of papers were found as 8.1, 14.4, and 40.6, respectively. The papers with 
corresponding author from Turkey constitute about 79\% of all papers. 

In 838 papers, the total number of contributed researchers from Turkey were found as 
290 professional astronomers and 183 of them hold PhD degrees. The distributions 
of PhD holders and co-researchers in the published papers were shown in Fig. 3. While 
the number of papers per year per individual PhD, range between 0.50-1.35, the number 
of papers per all researchers change between 0.40-0.95 per year. Average number of 
papers per PhD and all researches are 0.89 and 0.67 per year, respectively.

\begin{figure}
\centering
   \includegraphics[scale=0.57]{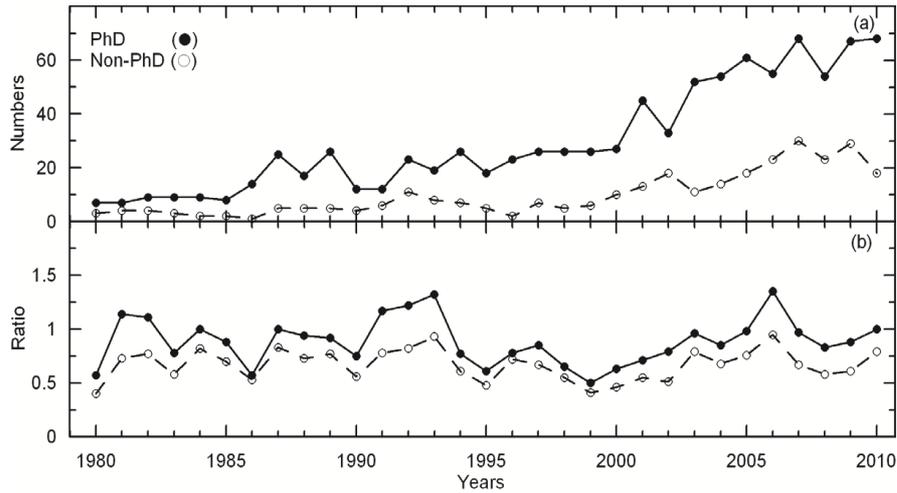}
   \caption{(a) The distribution of PhD and non-PhD holder researchers that 
   contributed to the papers published in SCI journals between 1980-2010 over 
   the years. (b) The ratio of papers to PhD and all of the researchers.}
\label{fields}
\end{figure}

One of the quality indicator of a scientific publication is the fact that if 
it can form a basis or have a supporting nature for prospective studies. This 
can be provided by receiving citations from other authors. In the last thirty 
years the number of citations to the identified 838 papers found to be 6938 
as of January 1, 2011. The average number of citations per paper is about 
8.3. If the number of self-citations subtracted from all of the citations, 
then the number of citations become 4559 and the average number of citations 
per paper become 5.4. In Fig. 4, the distribution of citations for total number 
of papers of each year were shown. It is easy to see from the figure that 
while the number of citations ascending linearly until the 2000s, there is a 
sudden jump afterwards. This jump can expressed in terms of exponential form: 
\begin{equation}
\ln N=0.1382\times t-270.9260   
\end{equation}
Here $N$ and $t$ are number of citations and years, respectively. The correlation 
coefficient $R^2$=0.94 indicates that the relation is well fit to the data. It is 
expected that the annual number of citations will exceed 1000 by the end of 2011. 
The analysis for 2011 showed that the number of citations for these papers 
is 995. From equation [1], the calculated number for 2011 was found 1042. This 
shows equation [1] is sensitive.

\begin{figure}
\centering
   \includegraphics[scale=0.40]{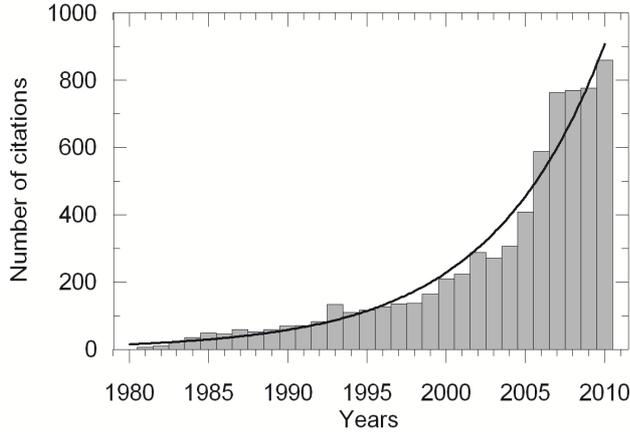}
   \caption{Distribution of citations of 838 papers that published in SCI 
    journals between 1980-2010.}
\label{fields}
\end{figure}

In three different periods between 1980 and 2010, the increase of the average 
number is found proportional to the number of researchers that contributed to 
these papers (Figs. 2 and 3). The parallel increase among the number of papers 
and researchers indicate that the number of papers per researcher do not 
increase significantly during the period 1980 - 2010. However, it was shown by
\cite{Abt07} that the publication rate depends on the number of scientists working
in natural sciences.
 
\subsection{The Distribution of Papers over Turkish Institutions}

To unveil the combined research performance of individual Turkish institutions, 
we selected papers with the lead author based in Turkey. We find that out of 
all 838 identified papers, 658 papers fall in this category (i.e., 79\% of all 
publications). In Table 1, the total number of papers originating from nine 
leading institutions, their citations and combined h-indices are listed. Note 
that COMU ({\c C}anakkale Onsekiz Mart University), Sabanc\i~University and 
Akdeniz University were established much later than others. Therefore, the 
year of the first publication of each institution in astronomy and astrophysics 
is also tabulated during the period 1980-2010. We find that the largest 
proportion of papers in the time frame from 1980 to 2010 originated from Ege 
University and METU (Middle East Technical University). There were 65 papers
originated from 24 institutions other than the nine listed here. Those 
institutes were combined in the last row of the table as the Others.
As far as the impacts of the papers are concerned, the number of citations per 
paper exceed 10 for three institutions: Bo{\u g}azi{\c c}i, Sabanc\i~and 
T{\"U}B{\.I}TAK. The institutional h-indices are quite similar to each other: 
majority ranging between 9 and 15, with an average value for the first eight in 
the list of 11.6.

\begin{table}
\setlength{\tabcolsep}{5pt}
\caption{Number of papers, their citations and h-index values of leading Turkish institutions.}
\begin{tabular}{clcccccc}
\hline
ID & University/ & Year of First &    Number & \% of  & Citations & Citations/paper & h-index \\
   &   Institute &   Publication &   of Paper&  658   &           &                 &         \\
\hline
 1 & Ege                  & 1980 & 116 & 17.63 & 646 &  5.57 & 11 \\
 2 & METU                 & 1980 & 112 & 17.02 & 902 &  8.05 & 15 \\
 3 & COMU                 & 1996 &  75 & 11.40 & 327 &  4.36 & ~9 \\
 4 & Bo{\u g}azi{\c c}i   & 1981 &  71 & 10.79 & 789 & 11.11 & 14 \\
 5 & {\.I}stanbul         & 1981 &  70 & 10.64 & 413 &  5.90 & 12 \\
 6 & Ankara               & 1980 &  64 &  9.73 & 315 &  4.92 & ~9 \\
 7 & T{\"U}B{\.I}TAK      & 1986 &  37 &  5.62 & 446 & 12.05 & 12 \\
 8 & Sabanc\i             & 2001 &  29 &  4.41 & 348 & 12.00 & 11 \\
 9 & Akdeniz              & 1993 &  19 &  2.89 &  47 &  2.47 & ~4 \\
10 & Others               & 1980 &  65 &  9.87 & 326 &  5.02 & 11 \\
\hline
\end{tabular}  
METU: Middle East Technical University, COMU: {\c C}anakkale Onsekiz Mart University\\ 
\end{table}

\subsection{The Distribution of Papers over Collaborated Countries}
The distribution of the identified 838 papers according to the countries of the 
collaborated groups is shown in Fig. 5. Most of the studies that been done in 
collaboration are with researchers from the United States of America (19.4\%). 
This is followed by Italy (10.5\%), Germany (6.8\%), England (5.5\%), Spain 
(4.6\%), France (4.0\%), Australia (3.1\%) and others (46.1\%), respectively 
(Fig. 5). It is prominently common that the first seven of the mostly collaborated 
countries are very advanced in astronomy and astrophysics.

\begin{figure*}
\centering
   \includegraphics[scale=0.60]{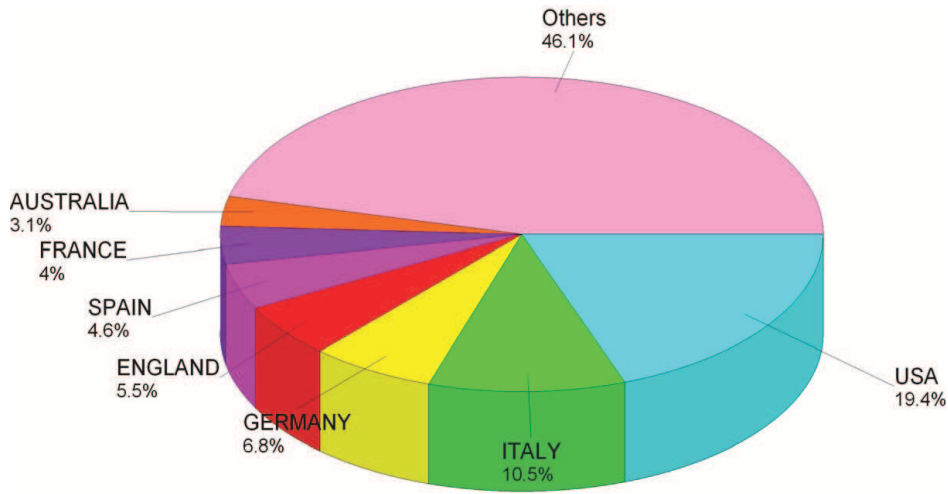}
   \caption{Distribution of collaborated countries from 1980 to 2010.}
\label{fields}
\end{figure*}

\subsection{The Journal Distribution of the Published Papers}
We also examined the 838 papers according to journal distribution in which they 
were published in Fig. 6. It was found that researchers from Turkey sent their studies 
mostly to the Astronomy and Astrophysics (21.1\%). This is followed by  
Astrophysics and Space Science (18.4\%), Monthly Notices of the Royal 
Astronomical Society (14.3\%), Astrophysical Journal (12.4\%), International 
Journal of Modern Physics D (6.9\%), New Astronomy (5.4\%), Astronomische 
Nachrichten (5.1\%), Astronomy and Astrophysics Supplement (2.7\%), Solar Physics 
(2.7\%), and others (11\%). If one considers the papers with the corresponding 
authors from Turkey (that is, 658 out of 838 papers), they were sent to 
Astrophysics and Space Science (22.2\%), Astronomy and Astrophysics (16.6)\%, 
Monthly Notices of the Royal Astronomical Society (15.3\%), Astrophysical Journal 
(8.8\%), International Journal of Modern Physics D (8.1\%), New Astronomy (6.5\%), 
Astronomische Nachrichten (6.1\%), Solar Physics (3.3\%), Astronomy and Astrophysics 
Supplement (2.9\%), and others (10.2\%). The distribution is shown in Fig. 7. 

\begin{figure}
\centering
   \includegraphics[scale=0.50]{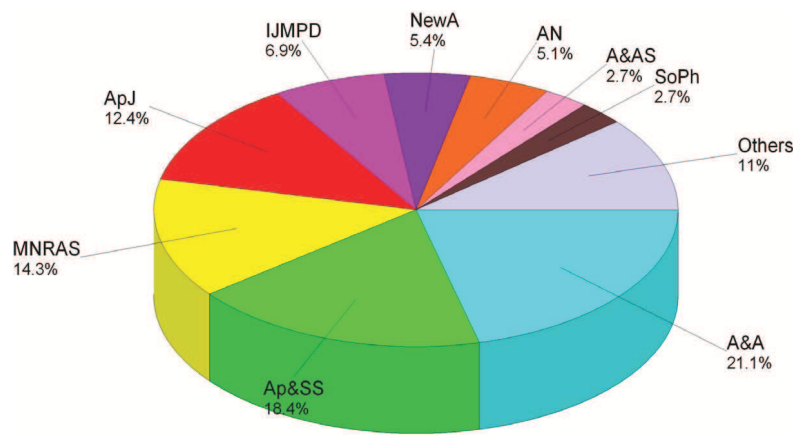}
   \caption{Distribution of published papers in SCI journals researchers from Turkey 
    from 1980 to 2010.}
\label{fields}
\end{figure}

\begin{figure}
\centering
   \includegraphics[scale=0.50]{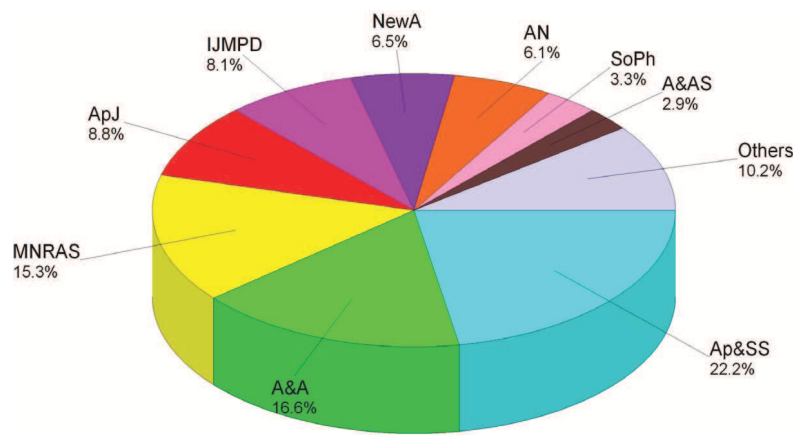}
   \caption{Distribution of published papers in SCI journals where as the corresponding 
   author from 1980 to 2010.}
\label{fields}
\end{figure}

By further investigating the SCI journals that preferred by the researchers from Turkey 
between 1980 and 2010, we find that seven journals constitute 84\% of all journal articles. 
The annual distribution of papers published in these seven journals are presented in Fig. 8. 
In the figure, filled and empty regions represent the rank of the authors depending on whether 
they are the corresponding author from Turkey, or not, respectively. When Astronomy and 
Astrophysics and Astrophysical Journal were taken into account, it is found that number of 
papers with corresponding authors from Turkey dropped significantly. The cause of such drop 
is clearly the page charges asked from Turkish authors by these two journals (Fig. 8). 
However, this situation is not the case for other journals, such as, Astrophysics and Space 
Science, Monthly Notices of the Royal Astronomical Society, International Journal of Modern 
Physics D, New Astronomy, Astronomische Nachrichten. Most of the papers published in these 
journals were made by the corresponding authors from Turkey. 

\begin{figure}
\centering
   \includegraphics[scale=0.55]{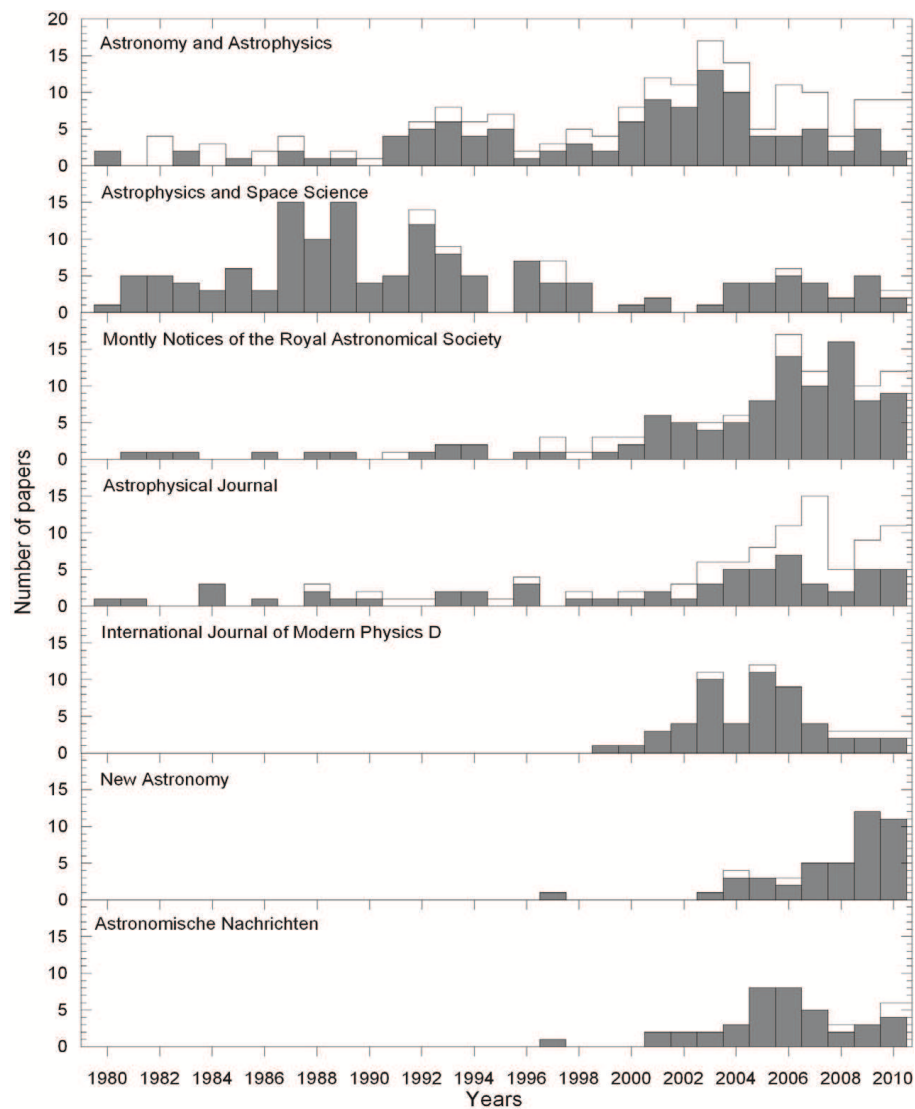}
   \caption{Distribution of papers in years over the seven SCI journals between 1980-2010 
    according to all authors (empty bars) and corresponding authors from Turkey (filled bars).}
\label{fields}
\end{figure}

\subsection{Impact of the Papers from 2006 to 2010}

Since impact factors of all journals included here are available only between 2006 and 2010, 
we limited our investigation of their impacts only in this range. There are 225 papers produced 
by researchers from Turkey in 2006 - 2010. This comprises of 27\% of all the published papers 
in the thirty year period. The distribution of papers based on their impact factor from 
2006 to 2010 is shown in the Fig. 9. In this period the papers that were sent and published in 
the SCI journals which have impact factors less than six. It is very clear from the Fig. 9 that 
the distribution has two peaks with median impact factor of three. The detailed analysis of the 
distribution shows that the average number of impact factors for less and more than three are 
1.65 and 4.97, respectively. It is clear that the astronomers from Turkish institutes publish 
equally in high and mid level journals. 

\begin{figure*}
\centering
   \includegraphics[scale=0.40]{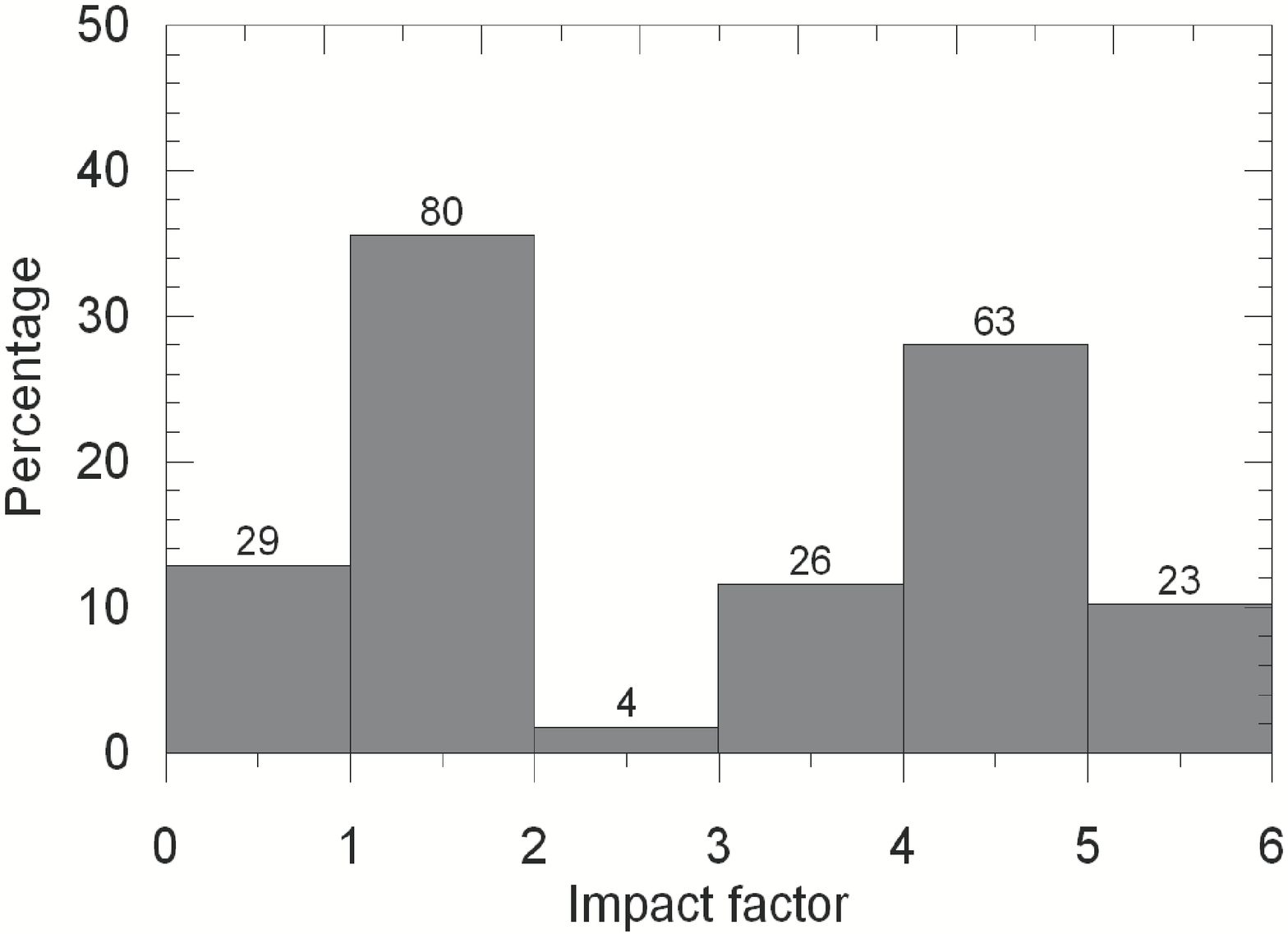}
   \caption{Distribution of 225 papers based on their impact factor of SCI journals in 2006-2010.}
\label{fields}
\end{figure*}

\subsection{Comparison of Research Performance with some OECD Countries}
The Organization for Economic Cooperation and Development\footnote{http://www.oecd.org} (OECD) 
is a body comprised of 30 developed and developing countries, aiming for developments through 
exchange of ideas. Here, we compared research performance of astronomers and astrophysicists 
in Turkey with those nations having similar Gross Domestic Expenditure (GDE) on Research and 
Development (R\&D). For this purpose, Chile, Mexico, Slovak Republic, Greece, Poland, Hungary 
and Italy are selected and the same search criteria explained earlier were applied. In 
Table 2, the number of publications in astronomy and astrophysics, total citations to these 
publications, their h-indices are listed for these seven selected OECD countries, as well as 
for Turkey.

\begin{table}
\setlength{\tabcolsep}{5pt}
\caption{Numbers of astronomy and astrophysics papers and their citations of comparison OECD countries from 1980 to 2010. This analysis were made as of November 2012.}
\begin{tabular}{lcccccccc}
\hline
Country                                   &      Chile &    Mexico  & Slovak     & Greece     & Poland     &    Turkey  & Hungary & Italy \\
                                          &            &            & Republic   &            &            &            &         &       \\
\hline
Number of papers                          &       5577 &       3494 &       367  &       1832 &       3734 &        838 & 1092    & 17882 \\
Sum of the times cited                    &     206184 &      84364 &      3816  &      30173 &     100450 &       7525 & 42804   & 567359\\
Sum of times cited without                &            &            &            &            &            &            &         &       \\
self-citations                            &     185388 &      72832 &      3416  &      26585 &      88000 &       6938 & 40471   & 503828\\
Citing articles                           &      84191 &      42057 &      2815  &      18712 &      51584 &       5755 & 24804   & 223839\\
Average citations per item                &      36.97 &      24.15 &     10.40  &      16.47 &      26.90 &       8.28 & 39.20   & 31.73 \\
h-index                                   &        146 &        102 &        26  &         70 &        115 &         37 & 83      & 219   \\
GDE on R\&D (*)                           &       0.36 &       0.39 &      0.51  &       0.59 &       0.61 &      0.69  & 1.02    & 1.17  \\
Number of IAU members                     &        106 &        124 &        42  &        109 &        160 &         46 & 56      & 587   \\
\hline
\end{tabular} 
(*) These indicate average percentage values of GDP between 2004-2010.\\
{\tiny
http://www.oecd-ilibrary.org/science-and-technology/gross-domestic-expenditure-on-r-d\_2075843x-table1\\
}
\end{table}

As seen in the Table 2, among these eight nations Italy has produced the highest number of 
publications and collected the highest number of citations in the period of 1980-2010. 
In both categories, Italy was followed by Chile, Poland, Mexico, Greece, Hungary, Turkey and 
Slovak Republic. In the category of number of citations per publication, Hungary is on top 
of these eight nations with 39.2 citations/paper, followed by Chile ($\sim$ 37 
citations/paper), Italy, Poland, Mexico, Greece, Slovak Republic and Turkey. In case of 
country h-index ranking, Italy again appears on top Chile, Poland, Mexico, Hungary, Greece, 
Turkey and Slovak Republic follow.

\section{Discussion and Conclusions}

We have investigated the research performance of scientists based in Turkey working in 
various fields of astronomy and astrophysics in the period of 1980-2010. As far as the 
annual number of papers are concerned, there are clear corner stones, around which 
publication productivity was increased remarkably. These two are around 1988 and 2000. 
The former one is probably a combined results of developments in computing and file 
sharing technologies, developments in space based observations and the launch of X-ray 
satellites (in particular EXOSAT), relatively easier access to journals via electronic 
submission, and the return of prominent scientists after the completion of their post 
graduate studies abroad. The other jump in productivity around 2000 is likely due to 
(i) initiation of the optical observations with the 1.5m Russian-Turkish telescope at 
the T\"UB\.ITAK national observatory {\footnote{http://www.tug.tubitak.gov.tr/}, (ii) 
graduation of a group of active researchers, who were attracted to astronomy and 
astrophysics in the second half of 1990s, (iii) participation in international 
collaborations, as well as (iv) return of scientists back to Turkey after receiving 
their PhD degrees and other factors mentioned above. There are other important factors 
that may have played role in the increase, such as, faster and efficient communication 
with international partners, obligation of producing SCI based scientific papers for 
academic promotions, and shortened evaluation periods via electronic handling and 
reviewing of the journal articles. 

The scientific impact of these publications (i.e., the number of citations) have been 
increasing rapidly. Note here that several shortcomings in our database searches
might have affected our sample (see a detailed discussion on bibliographic shortcomings
in \cite{Marx01} and  \cite{vanLeeuwen07}). The most probable shortcomings in evaluating 
the performance could originate from erroneous or inadequate citations, and insufficient 
time for an article to get citations after it is published. For the latter, we find that 
the average cited life time of journals investigated here is about six years. As our 
searches were performed on 2011 January 1, citation counts of the latest publications 
may not reflect their actual values.

The distribution of impact parameters of the refereed journal, to 
which Turkish astronomer and astrophysicists prefer to sent their manuscripts, shows a
bi-modality; the distribution has peaks around 1.7 and 5.0. Note that journals with 
impact factor larger than four usually require publishing charges while those with 
impact factor less than three are usually free of charge. As there is usually no 
available budget for page charge fees, many Turkish researchers prefer to send their 
papers to journals with no charge. Unavoidably, some of the higher quality studies may 
end up being published in journals with lower impact factor.

Among all astronomical and astrophysical study topics, papers on stellar astrophysics 
and high energy astrophysics constitute almost 80\% of all publications of Turkish 
scientists between 1980 and 2010. On the other hand, the fields of solar system 
astrophysics and astronomical instrumentation are severely under-studied. There has 
been a growing interest in the observational aspects of extra-solar planets but this 
extremely popular field has not been matured yet to yield scientific publications. 

Comparison of the scientific productivity of Turkish astronomers and astrophysicists 
with those in a set of OECD countries with similar R\&D spending from their gross
domestic products shows that Turkey is lagging behind both in number of publications 
and the number of citations per paper. It is not surprising that Italy leads in
many respects among these eight countries as it is the birth place of modern
astronomy (science). Chile's produces significantly large number of papers (which
are highly cited), even though its gross national expenditure on R\&D is the lowest
within the OECD countries. This is because Chile hosts some of the largest optical 
telescopes on its soil and attracts a lot of high profiles scientists as short or 
long term visitors. Another important factor that affects the productivity is the
number of active researchers. According to the International Astronomical Union (IAU)
records, Turkey has the second least number of IAU members (after Slovak Republic)
among all OECD countries. There had been other socioeconomic and cultural reasons behind 
the lag in productivity. Nevertheless, rapid growth in Turkish astronomy and 
astrophysics research over the recent years indicates that the gap is narrowing.

Developments in instrumental capabilities through domestic resources or international 
collaborations have been one of the main driving factor for astronomy and astrophysics 
in Turkey. The 1.5m Russian-Turkish Telescope and one of four Robotic Optical
Transient Search Experiment (ROTSE), both installed at the Bak\i rl\i tepe observing 
site, have impacted scientific productivity positively. These developments can be 
sustained by improving observing tools, building unique telescopes and joining large 
international bodies. Recently, Development Ministry of Turkey approved the foundation 
of a 4m class infrared telescope in the eastern part of Turkey 
{\footnote{http://dag-tr.org/}. In the future, with the increasing number of active 
researchers in Turkey, joining the European Southern Observatory (ESO) and European 
Space Agency (ESA) will likely be a necessity rather than a dream.

\begin{acknowledgements}
We thank the anonymous reviewer and Helmut A. Abt for insightful suggestions that have 
contributed to the improvement of the paper. We thank Cafer {\.I}bano{\u g}lu, Halil 
K\i rb\i y\i k, Osman Demircan, Zeki Eker for providing useful comments. This research 
has made use of Thompson Routers Web of Science.
\end{acknowledgements}

\end{document}